\documentclass[prb, preprint]{revtex4-1} 
\usepackage{amsmath} 
\usepackage{amsfonts} 
\usepackage[dvips]{graphicx}
\usepackage{amssymb}
\usepackage{amsmath}

\begin{document}
\title{The following article has been submitted to, and conditionally accepted by, the American Journal of Physics. After it is published, it will be found at http://scitation.aip.org/ajp/.
\bigskip
Spontaneous Symmetry Breakdown in non-relativistic Quantum Mechanics}
\author{R. Mu\~{n}oz-Vega}\email{rodrigo.munoz@uacm.edu.mx}
\affiliation{Universidad Aut\'{o}noma de la Ciudad de M\'{e}xico, Centro Hist\'{o}rico, Fray Servando Teresa de Mier 92, Col. Centro, Del. Cuauht\'{e}moc, M\'{e}xico D.F, C.P. 06080}
\author{A. Garc\'{i}a-Quiroz}\email{ e-mail:alberto.garcia@uacm.edu.mx}
\affiliation{Universidad Aut\'{o}noma de la Ciudad de M\'{e}xico, Centro Hist\'{o}rico, Fray Servando Teresa de Mier 92, Col. Centro, Del. Cuauht\'{e}moc, M\'{e}xico D.F, C.P. 06080}
\author{ Ernesto L\'{o}pez-Ch\'{a}vez}\email{elopezc-h@hotmail.com}
\affiliation{Universidad Aut\'{o}noma de la Ciudad de M\'{e}xico, Centro Hist\'{o}rico, Fray Servando Teresa de Mier 92, Col. Centro, Del. Cuauht\'{e}moc, M\'{e}xico D.F, C.P. 06080}
\author{Encarnaci\'{o}n Salinas-Hern\'{a}ndez}
\email{ esalinas@ipn.mx}
\affiliation{ESCOM-IPN, Av. Juan de Dios B\'{a}tiz s/n, Unidad Profesional Adolfo L\'{o}pez Mateos, Col. Lindavista, Del. G. A. Madero, M\'{e}xico, D. F, C.P. 07738}
\begin{abstract}
The  advantages and disadvantages of some pedagogical non-relativistic quantum-mechanical models, used to illustrate Spontaneous Symmetry Breakdown,  are discussed. A simple quantum-mechanical toy model (a spinor on the line, subject to a magnetostatic interaction) is presented, that exhibits the spontaneous breakdown of an internal symmetry. 
\end{abstract}
\date{\today}
\maketitle

\section{Introduction}
Whenever the ground state of a given physical system fails to exhibit a symmetry that is present in the fundamental equations of that system, it is said that this symmetry has \emph{spontaneously been broken.}

The spontaneous breakdown of a symmetry was first noticed in solid state physics and related fields, where it has played an important role in our understanding of phenomena such as superconductivity and ferromagnetism. Symmetry breakdown in that context has previously been addressed in education journals.\cite{Aravind}

In 1960 Y. Nambu \cite{Nambu1} offered the conjecture that some of the approximate symmetries observed in relativistic particle physics could be explained as spontaneously broken exact symmetries. The term \emph{Spontaneous Symmetry Breakdown} (SSB, from now on) was introduced the following year by M. Baker and S. L. Glashow.\cite{BakerGlashow}

An apparently unsurmountable objection to Nambu's conjecture was quickly raised by J. Goldstone, A. Salam and S. Weinberg. \cite{Goldstone} This objection, the Goldstone theorem, states that in every physically acceptable (i. e. covariant) theory, SSB brings with it the presence of unwanted massless particles (the so called Goldstone ghosts, or Goldstone bosons.)

This in turn lead to the proposition, by P. W. Anderson,\cite{Anderson} that the coupling of the system with a long-range field (such as the electromagnetic one) could remove the Goldstone ghosts from the theory.

Finally, in 1964 P. W. Higgs\cite{Higgs} proposed his celebrated mechanism, by means of which the Goldstone bosons are eliminated by coupling the currents associated with the broken symmetry with a gauge field. The search for traces of the Higgs mechanism then became one of the main obsessions of experimental and theoretical physicist, as is known by every newspaper reader.

There are many fine points we have left out of this very brief historical account. But at least one thing should come out clear: that SSB plays a central role in contemporary high-energy physics, so that it is only natural for introductory textbooks\cite{Halzen} to contain simplified SSB models. Indeed, the cited textbooks discuss a classical model: a point-like classical particle on the line, moving under the sole influence of a potential 
\begin{equation}\label{mexican.peasant.hat}
V(x)=\lambda x^{4}-\mu x^{2}
\end{equation} 
(with $\lambda$ and $\mu$ positive constants,) a typical example of a symmetric double well or \emph{sombrero}  potential. 

There are two different positions of stable equilibrium  (and thus, two different \emph{ground states}) in this model. Figure 1 makes this quite clear. In this example, the potential posses a global spatial-inversion, or $\mathcal{P}$-symmetry, i.e. the symmetry associated with the evenness of the potential about the axis $x=0$,
\begin{equation}
V(-x)=V(x)\quad \forall x\in\mathbb{R}\quad\textrm{ .}
\end{equation}
Because the equilibrium positions (each taken on its own) do not posses this symmetry, it is said that the $\mathcal{P}$-symmetry is violated by the (twice-fold degenerate) ground level of potential $V(x)$.
\begin{figure}[t]
\label{msbs1}
\begin{center}
\includegraphics[angle=0, width=\textwidth]{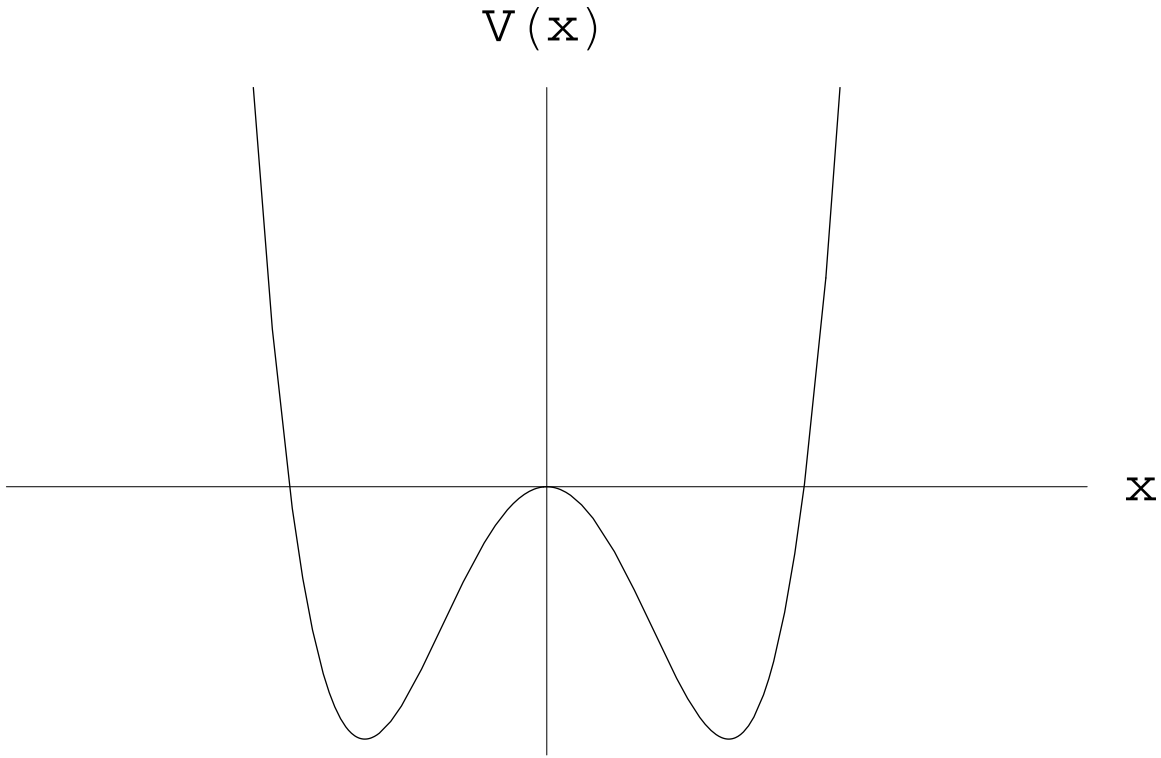}
\end{center}\caption{The potential $V(x)=x^{4}-x^{2}$,  an example of a \emph{sombrero} potential, with characteristic double minima and reflection symmetry. Conventional units.}
\end{figure} 

Thus, it is possible to illustrate SSB with classical models (other examples of  have been presented in this Journal.\cite{Sivardiere}) 
It is then natural to ask, first: If (non-relativistic) quantum-mechanical models can be constructed, that reflect the main features of SSB as they appear in Quantum Field Theories. And, in second term: if this purported non-relativistic quantum models offer any substantial pedagogical advantage over the classical ones. This article is addressed to answering this two questions. There is a previous answer in the well known textbook by Merzbacher, \cite{Merzbacher} which we comment and clarify in section IV.

This article has been written with senior and graduate students in mind. We think it may also be useful for faculty members, and general physicists, interested in obtaining a better grasp of its subject. The only requisite for understanding it is acquaintance with non-relativistic quantum mechanics (especially one-dimensional systems) in a level not above that of the first chapters of any university textbook on Quantum Mechanics, such as the one just mentioned.

The structure of the rest of this article is as follows: In section II we lay out the necessary notation and establish a lemma regarding the relation between SSB in one-dimensional Quantum Mechanics and the existence of non-overlapping symmetry-breaking pairs of ground eigenstates. In section III we establish that no one-dimensional continuous  potential with a lower bound can exhibit spontaneous symmetry breakdown, and study the consequences of this fact for a \emph{sombrero} potential. Section IV deals with models that \emph{do} exhibit spontaneous $\mathcal{P}$-symmetry breakdown, while section V is dedicated to the spontaneous breakdown of an internal symmetry.  Finally, conclusions are laid out in section VI.
\section{Spontaneous symmetry breakdown in non-relativistic quantum mechanics}
In Quantum Field Theories, SSB occurs only  for idealized systems that are infinitely large. This is so because in quantum systems the tunneling between ground eigenstates is possible. \cite{Weinberg} Obviously, this last feature is one that no classical model could ever reflect. In the following pages we will show how several different non-relativistic models can abet the understanding of the characteristics peculiar to \emph{quantum} SSB.

Let us start by laying out the fundamentals of SSB in terms of non-relativistic quantum theory. To this end, consider a quantum mechanical system described by a hamiltonian $\hat{H}$ endowed with a symmetry $\mathcal{U}$. This means that there is a symmetry operation, represented by a linear unitary operator\cite{dis1} $\hat{\mathcal{U}}$, such that:
\begin{equation}\label{con}
[\hat{H},\hat{\mathcal{U}}]=0\quad\quad\textrm{.}
\end{equation}  
Suppose now that there are a couple of normalized linearly independent vectors, $\vert A\rangle$ and $\vert B\rangle$ such that:
\begin{itemize}
\item[1)] $\quad\quad\vert A \rangle$ is an eigenvector of $\hat{H}$, and
\item[2)]  \begin{equation}\label{ssb}
\hat{\mathcal{U}}\vert A \rangle=\vert B\rangle\quad\textrm{ , }
\end{equation}
\end{itemize}
then it is easy to prove that  $\vert B\rangle$ is also an eigenvector of $\hat{H}$ with the same eigenvalue as $\vert A\rangle$, by using equation (\ref{con}). If two such eigenvectors exist for the ground energy level of $\hat{H}$, it is said that $\hat{H}$ spontaneously breaks symmetry $\mathcal{U}$. We shall call this couple an $\mathcal{U}$-symmetry-breaking pair.

Let us now state a simple lemma, that clarifies the relationship between SSB in one-dimensional quantum-mechanical systems, and the tunneling between eigenfunctions with the same eigenvalue: we assert that SSB occurs in one such  system iff the system is endowed with two orthogonal (i. e. non-overlapping) ground-level eigenstates that comply with (\ref{ssb}). The proof is as follows: suppose that there exists a couple of ground eigenstates, $\vert A\rangle$ and $\vert B\rangle$, that comply with (\ref{ssb}) such that:
\begin{equation}
\langle A\vert B\rangle\neq 0\quad \textrm{ . }
\end{equation} 
From the Cauchy-Schwarz inequality we have that:
\begin{equation}
\vert \langle A\vert B \rangle\vert <1\quad \quad \textrm{ , }
\end{equation}
so that normalized linearly independent eigenvectors $\vert +\rangle$ and $\vert -\rangle$, defined, respectively, by:
\begin{equation}\label{plus}
\vert +\rangle=\frac{1}{\sqrt{2(1+\langle A\vert B \rangle)}}\Big(\vert A \rangle +\vert B\rangle\Big)
\end{equation} 
and by:
\begin{equation}\label{minus}
\vert -\rangle=\frac{1}{\sqrt{2(1-\langle A\vert B\rangle)}}\Big(\vert A\rangle-\vert B\rangle\Big)
\end{equation} 
exist. (In one-dimension we can always choose the relative phase between $\vert A\rangle$ and $\vert B\rangle$ so that $\langle A\vert B\rangle$ is made real and 
\begin{equation}
\langle A\vert B\rangle=\langle B\vert A \rangle\quad \textrm{ , }
\end{equation}                
this just simplifies expressions (\ref{plus}) and (\ref{minus}), but it is not essential to the proof.)  
It can be checked, by applying the $\hat{\mathcal{U}}$ operator to this last expressions, that:
\begin{equation}
\hat{\mathcal{U}}\vert +\rangle=\vert +\rangle\quad\textrm{ , }\quad\hat{\mathcal{U}}\vert -\rangle=-\vert -\rangle\quad\textrm{ and }\quad\langle +\vert -\rangle=0\quad\textrm{,}
\end{equation}
so that $\vert+\rangle$ and $ \vert -\rangle$ are the elements of an orthonormal  basis for the subspace spanned by $\vert A\rangle$ and $\vert B\rangle$. We now define the pair formed by vectors:
\begin{equation}
\vert L\rangle\ = \ \frac{1}{\sqrt{2}}(\vert +\rangle+\vert -\rangle)\quad\textrm{ and }
\quad \vert R\rangle=\frac{1}{\sqrt{2}}(\vert +\rangle-\vert -\rangle)\quad\textrm{ .}
\end{equation}
This pair complies with:
\begin{equation}
\hat{\mathcal{U}}\vert L\rangle=\vert R\rangle
\end{equation}
and with:
\begin{equation}
\langle L\vert R\rangle=0 \quad\quad\textrm{.}
\end{equation}
Each one of the preceding steps can be checked by hand from the definitions. This completes the proof. In so many words: a  one-dimensional system exhibits SSB iff there is a null transition probability between a pair of symmetry-breaking ground level eigenstates, so that there is no tunneling between these eigenstates. We will discuss some of the consequences of this lemma in the next two sections.

As subproducts, we have two curious sub-lemmas: first, that if a symmetry is broken for a one-dimensional system then it has (at least) the eigenvalues $\pm 1$. And in second term, that one can always form a pair of symmetry-respecting non-overlapping eigenvectors, $\vert +\rangle$ and $\vert -\rangle$.
\section{\emph{Sombrero}-type potentials in one-dimensional non-relativistic Quantum Mechanics}
A theorem of elementary Quantum Mechanics tells us that the bounded part of the spectrum of a one-dimensional hamiltonian with lower bounded continuous potential is non-degenerate.\cite{LLandau} Thus, no one-dimensional continuous lower bounded potential (\emph{sombrero} or otherwise) can exhibit SSB. Period. One-dimensional potentials with broken symmetry must be discontinuous. And not just any discontinuity will do: it has to be \emph{strong enough} (so to speak) as to prevent the overlapping of a symmetry-breaking pair of ground level eigenstates. As we will learn in section IV, an insurmountable central barrier does the trick for $\mathcal{P}$-symmetry breaking.

Consequence of all the above, a quantum \emph{sombrero} potential acts not as classical \emph{sombrero} potential, but instead exhibits the usual quantum behavior: non-degenerate spectrum, states classified according with parity, number of nodes increasing with energy, etc. Let us study an illustrative example that may help dispel some unconsciously carried misconceptions on the subject. To this end, consider a function, $f:\mathbb{R}\rightarrow \mathbb{R}$, given by:
\begin{equation}
f(x)=e^{-ax^{4}}
\end{equation}
for some $a>0$. This function is clearly normalizable:
\begin{equation}
\int_{-\infty}^{\infty}\vert f (x)\vert^{2}dx=\int_{-\infty}^{\infty}e^{-2ax^{4}}dx<\infty\quad\textrm{.}
\end{equation}
The operator $\hat{q}$, defined by:
\begin{equation}\label{anh}
\hat{q}=-\imath\frac{d}{dx}-\imath 4ax^{3}
\end{equation}
annihilates $f$, \emph{i.e.}
\begin{equation}
\hat{q} f(x)=0\quad\textrm{,}
\end{equation}
\begin{figure}[t]
\label{msbs4_3}
\begin{center}
\includegraphics[angle=0, width=\textwidth]{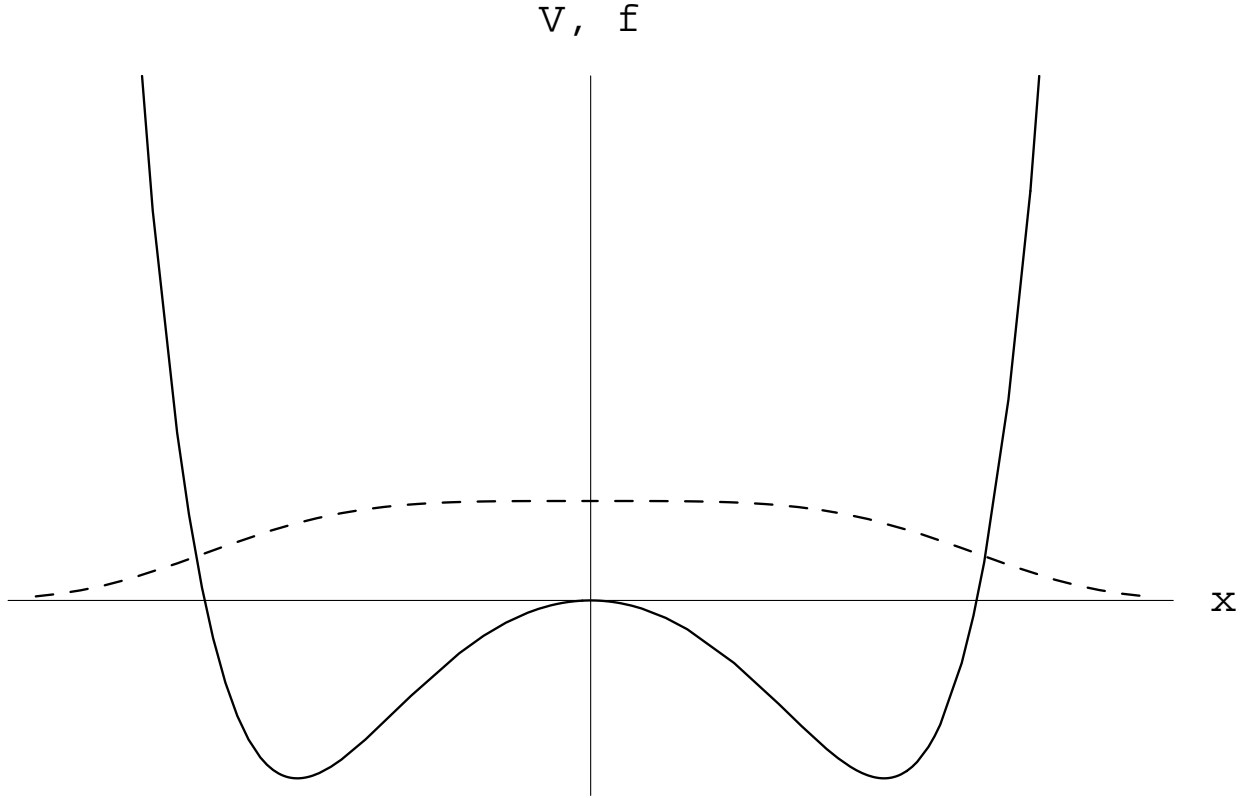}
\end{center}
\caption{The potential $V(x)$ of equation (21) (shown in black) along with $f(x)$ (shown dashed.) Conventional units.} 
\end{figure}
so that the explicitly self-adjoint operator $\eta$, defined by:
\begin{equation}
\hat{\eta}=\hat{q}^{\dagger}\hat{q}
\end{equation}
also annihilates $f$. The action of $\hat{\eta}$ on any given function, $\psi (x)$, is easily calculated:
\begin{equation}
\hat{\eta}\psi (x)=(-\imath\frac{d}{dx}+\imath 4ax^{3})(-\imath\frac{d}{dx}-\imath 4ax^{3})\psi (x)=\Big(-\frac{d^{2}}{dx^{2}}+16a^{2}x^{6}-12ax^{2}\Big)\psi (x)
\end{equation}
Thus, for any value $m>0$ we can construct a hamiltonian, $\hat{H}$, of the typical Schroedinger form:
\begin{equation}\label{herohat}
\hat{H}=\frac{\hbar^{2}}{2m}\hat{\eta}=\frac{\hat{p}^{2}}{2m}+V(x)
\end{equation} 
with a potential $V(x)$ given by:
\begin{equation}\label{pottyhat}
V(x)=\frac{\hbar^{2}}{2m}\Big(16a^{2}x^{6}-12ax^{2}\Big)\quad\textrm{.}
\end{equation}
Because $f$ is also node-less we know $f$ to be proportional to the ground state eigenfunction of $\hat{H}$. Thus, the ground level energy is zero:
\begin{equation}
\hat{H}\phi=0\quad\textrm{.}
\end{equation}
where $\phi (x)$ is given by:
\begin{equation}
\phi(x)=\bigg(\int_{-\infty}^{\infty}\vert f (x)\vert^{2}dx\bigg)^{-1}f(x)\quad\textrm{ .}
\end{equation}
Potential (\ref{pottyhat}) is of the \emph{sombrero} type. Figure 2 shows it along with $f(x)$. Observe that in this example the energy eigenvalue of the ground-state (\emph{the most stable stationary state!}) corresponds to the energy level of the separatrix  curve of the analogous classical system. Moreover, the probability density of $ \phi$ has its peak, and is centered, in what would be the position of unstable equilibrium in the classical counterpart of the potential.
\section{Potentials with spontaneous $\mathcal{P}$-symmetry breakdown}
In his well known textbook, E. Merzbacher\cite{Merzbacher} provides us with an heuristic SSB model based on a potential taken from molecular physics. It starts with a family of double oscillator potentials given by the expression: 
\begin{equation}
V_{a}(x)=m\omega^{2}(\vert x\vert-a)^{2}\quad\textrm { , }\quad a>0\quad\textrm{ .}
\end{equation}
Figure 3 shows a typical member of the of this family of potentials. Merzbacher asserts that, as the limit $a\rightarrow \infty$ is approached ``\ldots two degenerate ground state wave functions are concentrated in the separate wells and do not have definite parity. Thus, the reflection symmetry\ldots is said to be hidden or broken spontaneously\ldots" It is cogent that Merzbacher's ``concentrated'' ground  functions  (let us call them $\psi^{\pm \infty}$) comply with what we have established as the \emph{sine qua non} conditions of quantum SSB:
\begin{figure}[t]
\label{cochise}
\begin{center}
\includegraphics[angle=0, width=\textwidth]{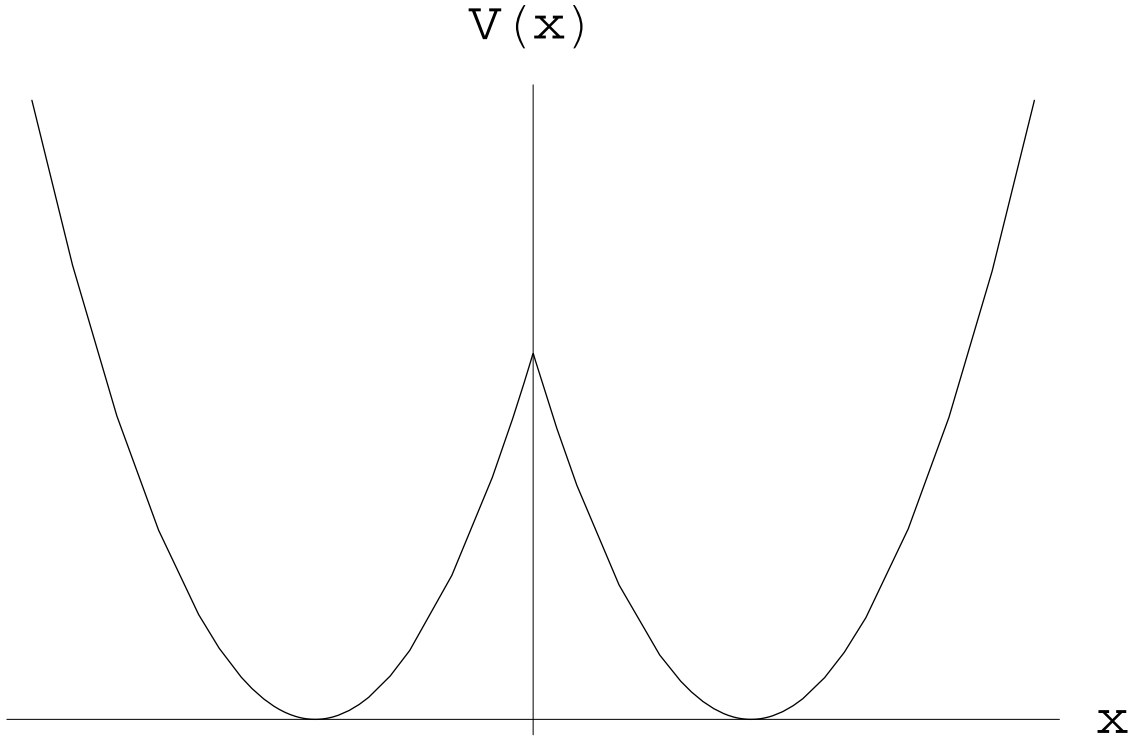}
\end{center}
\caption{An example of the double-oscillator potentials $V_{a}(x)$. Adapted from the textbook by E. Merzbacher.}
\end{figure}
\begin{equation}
\langle \psi^{+\infty}\vert\psi^{-\infty}\rangle=0
\end{equation}
and:
\begin{equation}
\hat{\mathcal{P}}\psi^{+\infty}=\psi^{-\infty}\quad\textrm{ , }
\end{equation}
where $\hat{\mathcal{P}}$ is the parity operator, defined by its action on any given function $g(x)$:
\begin{equation}
\hat{\mathcal{P}}g(x)=g(-x)\quad \quad\textrm{ . }
\end{equation}
In other words: Merzbacher's ``concentrated'' states constitute a $\hat{\mathcal{P}}$-symmetry-breaking pair of non-overlapping ground eigenstates, so that the model does indeed present SSB.

In order to study some of the aspects of  Merzbacher's model, we have constructed a thoroughly workable example of a $\mathcal{P}$-symmetry-breaking potential. This example starts with piece-wise-constant potentials of the type:
\begin{equation}\label{ucases}
U_{\alpha}(x)=\left \{ \begin{array}{c l}
\infty &\textrm{for } x \leq -a\\
0 &\textrm{if } -b > x > -a\\
\alpha &\textrm{for} b\geq x \geq -b\\
0 &\textrm{if } a> x > b\\
\infty &\textrm{if } x \geq a\\
\end{array}\right .\quad\quad\textrm{.}
\end{equation}
Figure 4 shows a typical member of this family. In the limit when $\alpha \rightarrow\infty$, one gets two infinitely deep square wells separated by a finite distance:
\begin{equation}\label{uinfty}
U_{\infty}(x)=\left \{ \begin{array}{c l}
\infty &\textrm{if } x \leq -a\\
0 &\textrm{if } -b > x > -a\\
\infty &\textrm{if } b\geq x \geq -b\\
0 &\textrm{if } a> x > b\\
\infty &\textrm{if } x \geq a\\
\end{array}\right .\quad\quad\textrm{.}
\end{equation}
Figure 5 shows this limit potential. Each $U_{\alpha}$ ($0<\alpha<\infty$) has a completely discrete spectrum, classified according to parity, with an infinite number of levels above $E=\alpha$. Levels start appearing below the barrier after some threshold value $\alpha_{0}$ is reached in the parameter. Let us focus in the discretization condition below barrier $E=\alpha$. For even states  it reads: 
\begin{equation}
E\cot^{2}(a-b)\frac{\sqrt{2mE}}{\hbar}=(\alpha-E)\tanh^{2}b\frac{\sqrt{2m(\alpha-E)}}{\hbar}\quad\textrm{,}
\end{equation}
\begin{figure}[t]
\label{betty}
\begin{center}
\includegraphics[angle=0, width=\textwidth]{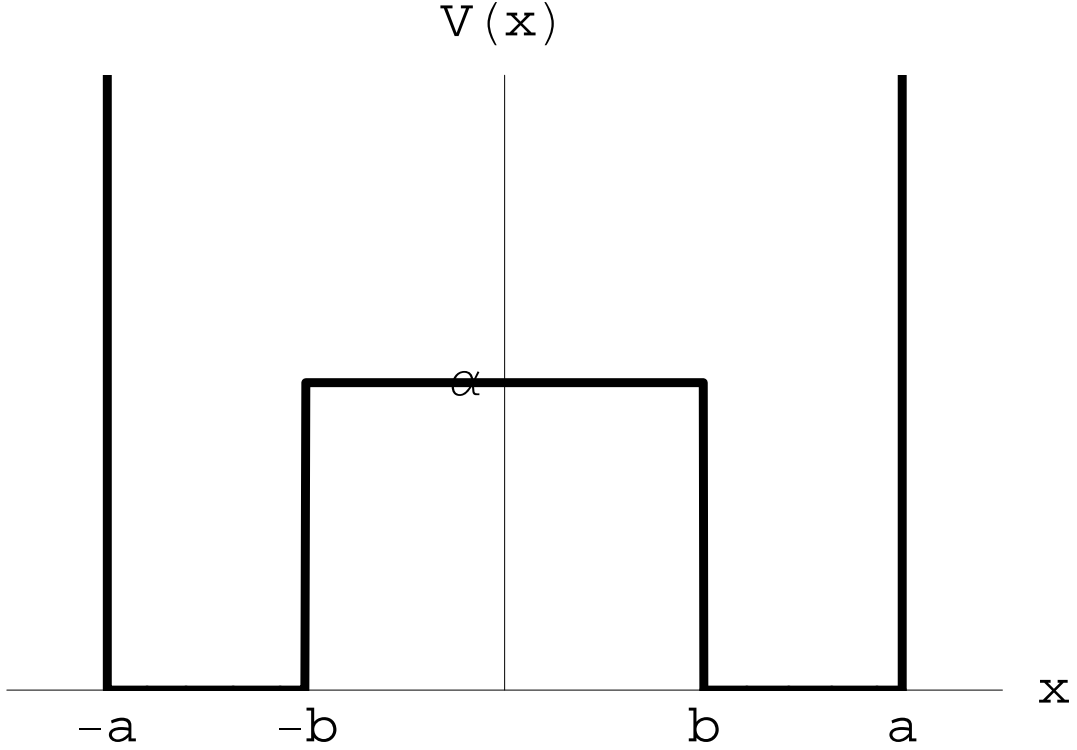}
\end{center}
\caption{An example of the potentials $U_{\alpha}(x)$ of equation (\ref{ucases}).}
\end{figure}
while odd levels below the barrier level have to comply with:
\begin{equation}
E\cot^{2}(a-b)\frac{\sqrt{2mE}}{\hbar}=(\alpha-E)\tanh^{-2}b\frac{\sqrt{2m(\alpha-E)}}{\hbar}\quad\textrm{.}
\end{equation} 
In the limit $\alpha\rightarrow\infty$ both of this expressions diverge to:
\begin{equation}
E\cot^{2}(a-b)\frac{\sqrt{2mE}}{\hbar}=\infty\quad\textrm{,}
\end{equation}
and this only makes sense if:
\begin{equation}
E_{n}=\frac{\pi^{2}\hbar^{2} n^{2}}{2m(a-b)^{2}}\quad\textrm{ , } \quad n=1,2,\ldots\quad\textrm{,}
\end{equation}
which is the usual discretization condition for a single infinitely deep well of width $a-b$. Thus, for the limit potential $U_{\infty}$, even and odd levels merge, so that the each level is twice-fold degenerate.

Symmetry-breaking pairs of the non-overlapping eigensolutions, $\psi_{L,n}$  and $\psi_{R,n}\ $, appear for every energy level of the stationary Schroedinger equation:
\begin{equation}
\Big\{-\frac{\hbar^{2}}{2m}\frac{d^{2}}{dx^{2}}+U_{\infty}(x)\Big\}\psi_{j,n}=E_{n}\psi_{j,n} \quad j=L,R\quad\textrm{,}
\end{equation}
This pairs are given by:
\begin{equation}\label{lefty}
\psi_{L,n}(x)=\left\{\begin{array}{cc}
\sqrt{\frac{2}{a-b}}\ \sin \frac{\pi n (x+a)}{a-b}& \textrm{ if } x\in(-a,-b)\\
 & \\
0&\textrm{elsewhere}
\end{array}\right.
\end{equation}
\begin{figure}[t]
\label{sussy}
\begin{center}
\includegraphics[angle=0, width=\textwidth]{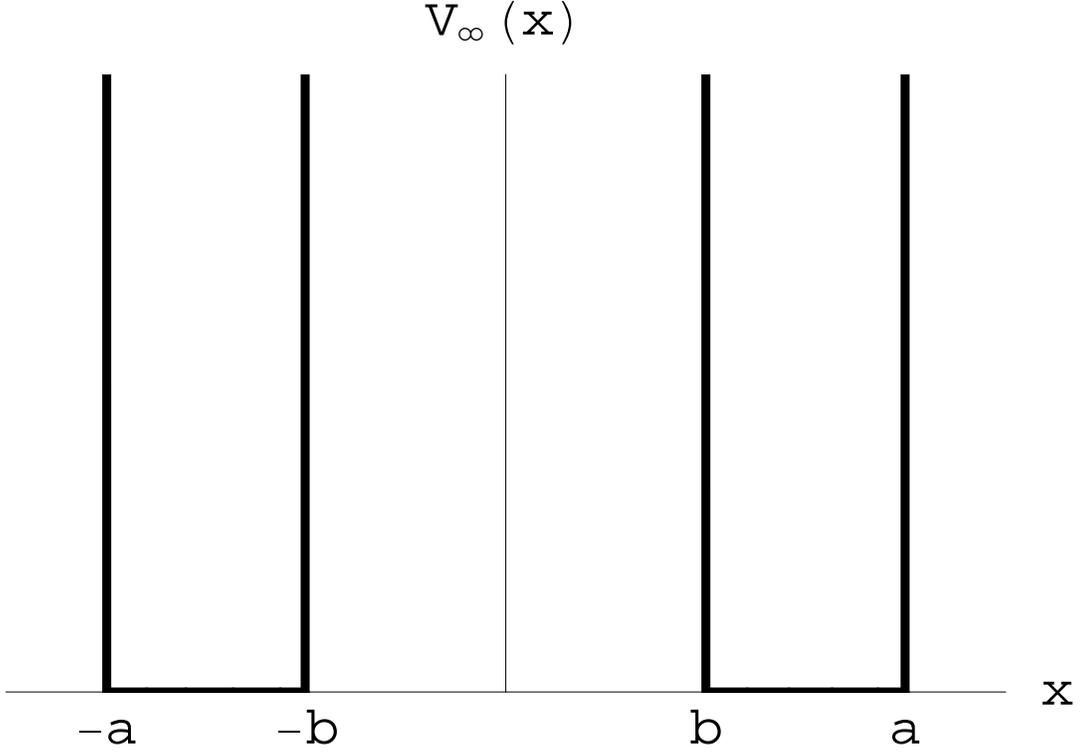}
\end{center}
\caption{ The $U_{\infty}$ potential: a double inifinite square well.}
\end{figure} 
and
\begin{equation}\label{righteous}
\psi_{R,n}(x)=\left\{\begin{array}{cc}
\sqrt{\frac{2}{a-b}}\ \sin \frac{\pi n (x-a)}{a-b}& \textrm{ if } x\in(b,a)\\
 & \\
0&\textrm{elsewhere}
\end{array}\right.\quad\quad\textrm{.}
\end{equation}
It is easy to check by hand that:
\begin{equation}
\langle \psi_{R,n}\vert\psi_{L,n}\rangle=0\quad n=1,2\ldots\quad \textrm{,}
\end{equation}
and that:
\begin{equation}
\hat{\mathcal{P}}\psi_{L,n}(x)=\psi_{R,n}(x) \quad n=1,2,\ldots\quad\textrm{,}
\end{equation}
As we shall see in a moment, this feature (the existence of symmetry-breaking non-overlapping pairs in each level) is shared by all $\mathcal{P}$-symmetric potentials with a central infinite square barrier.

There are a couple more of consequences  that we can draw from this example. First, as noticed in Section II, we can always find parity-respecting eigenfunctions:
\begin{equation}\label{evenodd}
\psi_{+,n}=\frac{1}{\sqrt{2}}\Big(\psi_{L,n}+\psi_{R,n}\Big)\quad\textrm{ , }\quad\psi_{-,n}=\frac{1}{\sqrt{2}}\Big(\psi_{L,n}-\psi_{R,n}\Big)\quad\textrm{.}
\end{equation}
The breakdown of symmetry brings with it a degeneracy that \emph{allows one to prepare the system in definite-parity eigenstates, among other possible states. The telling symptom of quantum-mechanical SSB is the existence of symmetry-breaking non-overlapping pairs such as} (\ref{lefty}-\ref{righteous}) \emph{and not the impossibility of finding even-odd pairs such as} (\ref{evenodd}). Moreover, the separation between the wells has nothing to do with the breaking of symmetry. In our example this separation is arbitrary. Changing the distance between the wells, or even taking it to infinity, will not change the behavior of the solutions.

Finally, let us show that every $\mathcal{P}$-symmetric potential endowed with a central square barrier of infinite height, breaks symmetry in each one of its levels (excited as well as ground level). In order to do this just consider a normalized solution, $\Psi$ for a given energy level of a $\mathcal{P}$-symmetric potential with a infinite barrier of width $L$ around $x=0$. Then $\Psi_{R}$, given by
\begin{equation}\label{limbaugh}
\Psi_{R}(x)=kh(x)\Psi(x)
\end{equation}  
will also be a solution for that same level, given that $k$ is the normalization factor:
\begin{equation}\label{k}
k\equiv \Big(\ \int_{L/2}^{\infty}dx \vert\Psi(x)\vert^{2}\ \Big)^{-1}
\end{equation}
and $h$ is the step-function:
\begin{equation}
h(x)\equiv \left\{\begin{array}{c c}
0&\textrm{ if } x<0\\
1&\textrm{ if } x\geq 0\\
\end{array}\right .
\end{equation}
Linearly independent from $\Psi_{R}$ there is another solution, $\Psi_{L}$ given by 
\begin{equation}
\Psi_{L}=\hat{\mathcal{P}} \Psi_{R}
\end{equation}
Even more, $\Psi_{L}$ and $\Psi_{R}$ are non overlapping. The only \emph{if} in this proof is the possibility that $h(x)\Psi(x)$ may be identically zero, so that expression (\ref{k}) makes no sense.  In this later case, $\Psi$ and $\hat{\mathcal{P}}\Psi$ constitute themselves a non-overlapping pair. This completes the proof (we have left out some details for the reader to fill.) 
\section{An example of a spontaneously broken internal symmetry}
Are there any one-dimensional non-relativistic quantum-mechanical models of SSB that do not present degenerate excited levels? The answer we have found is bittersweet, as it demands the introduction of an internal degree of freedom, such as spin. The good news is that this lifts the ban on continuous potentials.  Indeed, consider the following pair of  energy displaced harmonic oscillators:
\begin{eqnarray}\label{twice}
\hat{H}_{+}=-\frac{\hbar^{2}}{2m}\frac{d^{2}}{dx^{2}}+\frac{m\omega_{+}^{2}}{2}x^{2}-\frac{\hbar \omega_{+}}{2}\nonumber\\
 \nonumber\\
\hat{H}_{-}=-\frac{\hbar^{2}}{2m}\frac{d^{2}}{dx^{2}}+\frac{m\omega_{-}^{2}}{2}x^{2}-\frac{\hbar \omega_{-}}{2}\nonumber\\
\quad\textrm{.}\end{eqnarray}
and take incommensurable fundamental frequencies, \emph{i.e.} take frequencies $\omega_{\pm}>0$ such that the quotient: $\omega_{+}/\omega_{-}$ is irrational:
\begin{equation}
\label{irrational}
\frac{\omega_{+}}{\omega_{-}}\in\mathbb{R}-\mathbb{Q}\quad\quad\textrm{ .}
\end{equation}
Let us now define the matrix arrangement $\mathbb{H}$ as:
 \begin{equation}
\mathbb{H}=
\left(\begin{array}{cc}
\hat{H}_{+}&0\\
0&\hat{H}_{-}\\
\end{array}
\right)\quad\textrm{,}
\end{equation}
which is to act on spinors, \textrm{i.e.} arrangements of the form:
\begin{equation}
\Psi=\left(\begin{array}{c}
\psi_{+}\\
\psi_{-}\\
\end{array}\right)\qquad\textrm{ . }
\end{equation}
Then, $\mathbb{H}$ has eigenvalues and eigenspinors given by equation:
\begin{equation}
\mathbb{H}\Psi_{\pm,n}=\hbar n\omega_{\pm}\Psi_{\pm,n}
\end{equation}
where 
\begin{equation}
\Psi_{+,n}=\left(\begin{array}{c}
\psi_{+,n}\\
0\\
\end{array}\right)\quad\textrm{ , }\quad 
\Psi_{-,n}=\left(\begin{array}{c}
0\\
\psi_{-,n}\\
\end{array}\right)\quad\textrm{,}
 \end{equation}
the $\psi_{\pm,n}$ ($n\in\mathbb{N}\bigcup \{0\}$) standing for  the well known eigenfunctions of hamiltonians $H_{\pm}$ (the Hermite polynomials.)

The Pauli matrix
\begin{equation}
\sigma_{3}=\left(\begin{array}{cc}
1&0\\
0&-1\\
\end{array}\right)
\end{equation}
which is self-adjoint and unitary, i.e.
\begin{equation}
\sigma_{3}^{\dagger} \sigma_{3}=\sigma_{3}^{2}=\mathbb{I}\quad\textrm{,}
\end{equation}
($\mathbb{I}$ standing for the $2\times 2$ identity matrix) is an internal symmetry of the system:
\begin{equation}
[\sigma_{3},\mathbb{H}]=0
\end{equation}
and eigenspinors are classified according with this symmetry:
\begin{equation}
\sigma_{3}\Psi_{\pm,n}=\pm\Psi_{\pm,n}\quad\textrm{.}
\end{equation}
Condition (\ref{irrational}) guarantees that there are no  degenerate excited levels in the spectrum of $\mathbb{H}$. Yet, the ground state has a twice-fold degeneracy:
\begin{equation}
\mathbb{H}\Psi_{\pm,0}=0\quad\textrm{.}
\end{equation}
Thus, in this model symmetry is broken only at ground state level. The non-overlapping symmetry-breaking pair is given by:
\begin{equation}
\Psi_{R}=\frac{1}{\sqrt{2}}\left(\begin{array}{c}
\psi_{+,0}\\
\psi_{-,0}\\
\end{array}\right)\quad\textrm{ , }\quad \Psi_{L}=\sigma_{3}\Psi_{R}=\frac{1}{\sqrt{2}}\left(\begin{array}{c}
\psi_{+,0}\\
-\psi_{-,0}\\
\end{array}\right)
\end{equation}
This example may be given a little more \emph{physical} (in contrast with purely mathematical) appearance by writing $\mathbb{H}$ in the form:
\begin{equation}\label{model}
\mathbb{H}=\hat{H}_{0}\mathbb{I}-\frac{\hbar}{2}\mathbf{B}\cdot\sigma\quad \textrm{,}
\end{equation}
where $\hat{H}_{0}$ is just an energy displaced harmonic oscillator:
\begin{equation}
\hat{H}_{0}=-\frac{\hbar^{2}}{2m}\frac{d^{2}}{dx^{2}}+\frac{m\omega_{0}^{2}}{2}x^{2}-\epsilon_{0}\quad \textrm{,}
\end{equation}
$\mathbf{B}(x)$, given by:
\begin{equation}
B_{x}\equiv 0\quad \textrm{ , }B_{y}\equiv 0\quad\textrm{ , }B_{z}(x)= -\frac{2}{\hbar}\Big(\frac{m \omega_{\Delta}^{2}}{2}x^{2}-\epsilon_{\Delta}\Big)\quad\textrm{.}
\end{equation}
is a magnetostatic field and $\sigma=(\sigma_{1},\sigma_{2},\sigma_{3})$ has the Pauli matrices as components. In order to do this, one just needs to define:
\begin{equation}
 \begin{array}{cc}
\omega_{0}=\sqrt{\frac{\omega_{+}^{2}+\omega_{-}^{2}}{2}}\quad\textrm{ , }& \epsilon_{0}=\frac{\hbar (\omega_{+}+\omega_{-})}{2}\\
 & \\
\omega_{\Delta}=\sqrt{\frac{\omega_{+}^{2}-\omega_{-}^{2}}{2}}\quad\textrm{ , }&\epsilon_{\Delta}=\frac{\hbar (\omega_{+}-\omega_{-})}{2}\\
\end{array}\quad\textrm{.}
\end{equation}
Note that this example is somewhat related to the Zeeman effect: if one ``turns off'' the magnetostatic field $\mathbf{B}$ in (\ref{model}), all levels become degenerate. In our model $\mathbf{B}$ lifts the degeneracy only on the excited levels, leaving the ground level twice-fold degenerate.

 So: Is this SSB at all? It is certainly not the SSB of fundamental interactions that high-energy physicist are looking for, yet it complies with all the formal requirements. And it is not entirely without precedent. The Mottelson-Bohr model of deformed nuclei presents an example of SSB, well known to researchers in that field. \cite{Heyde, Landau2} One of the relevant quantum numbers for SSB in the Mottelson-Bohr model, although not an internal degree of freedom, has some of the properties of an angular-moment, just as spin. Also, there has been some research on isospin (internal) symmetry breakdown done over the years,\cite{isospin} although the resemblance of this later instance with  the example here presented should be taken with extreme caution.

In any case, we believe to have shown in this section that SSB  is a widely applicable concept, that goes well beyond $\mathcal{P}$-symmetry breakdown, so that the geometric reasoning so prominent in first presentations of the subject may actually be misleading.\cite{W}        

\section{Conclusions}
In the preceding pages we have called attention to the following facts:
\begin{itemize}
\item[-]That there are features of QFT-SSB that no classical model can reflect, but that can be illustrated with simple non-relativistic quantum models,
\item[-]that in non-relativistic one-dimensional Quantum Mechanics, SSB necessarily implies the existence of a pair of \emph{non-overlapping} symmetry-breaking ground-level eigenstates, 
\item[-] that this later condition is analogous to one in Quantum Field Theory, \emph{viz.} that field-theoretic SSB occurs only in idealized, infinitely large systems,   
\item[-]that there is no exact SSB for structureless particles on the line subject to lower bounded continuous potentials,
\item[-]that one-dimensional even potentials with an square infinite central barrier exhibit symmetry breakdown in all levels (ground as well as excited), , and finally
\item[-]that there are viable, and possibly pedagogically relevant, simple quantum-mechanical models with spontaneously broken internal symmetries.
\end{itemize}
\section{Acknowledgments}
The support of SNI-CONACYT (Mexico) is duly acknowledged. All authors gratefully recognize that the comments of anonymous referees have greatly improved the content of this article. E. S. H. also acknowledges the support of IPN and EDI-IPN (Mexico).

\textbf{Figure captions.}

Figure 1.  The potential $V(x)=x^{4}-x^{2}$, an example of a \emph{sombrero} potential, with characteristic double minima and reflection symmetry. 

Figure 2.The potential $V(x)$ of equation (21) (shown in black) along with $f(x)$ (shown dashed.) Conventional units.

Figure 3. An example of a double oscillator potential $V_{\alpha}(x)$. Adapted from the textbook by E. Merzbacher.

Figure 4. From equation (27), an example of the $U_{\alpha}(x)$ potentials. 

Figure 5. The $U_{\infty}$ potential: a double inifinite square well.

\begin{thebibliography}{99}
\bibitem{dis1} We are leaving out of this discussion the anti-linear anti-unitary time-reversal operator, $\mathcal{T}$ and the products $\mathcal{PT}$ and $\mathcal{CPT}$ ($\mathcal{C}$ being the charge operator) which are very relevant for SSB in QFT.
\end{thebibliography}
\end{document}